# Low-T Thermo: a new program for arbitrarily combining low-T thermochronological data to model thermal history

Ruxin Ding[a, b, c,]*

[a] School of Earth Science and Engineering, Sun Yat-sen University, Guangzhou 510275, China
[b] Guangdong Provincial Key Laboratory of Mineral Resources & Geological Processes, Guangzhou 510275, China
[c] Low-T Lab, Shanghai 201101, China

*Corresponding author: dingrux@mail.sysu.edu.cn

**Abstract**: A robust code, called Low-T Thermo, has been developed to combine low-T thermochronological data arbitrarily to model thermal history. After apatite fission-track age and confined length are decoupled into two completely independent data to inverse thermal history and thermal history inversion using mica Ar-Ar age or bedrock quartz optically stimulated luminescence age are developed, there are eight kinds of low-T thermochronological data used to inverse thermal history including apatite fission-track age, apatite fission-track confined length, zircon fission-track age, apatite (U–Th)/He age, zircon (U–Th)/He age, mica Ar-Ar, bedrock quartz optically stimulated luminescence age and vitrinite reflectance. A total of 247 kinds of combination modes can be used to jointly inverse thermal history in theory (except the eight single methods modelling). These arbitrary combinations are helpful to model thermal history with the "incomplete" low-T thermochronological data set regarded to be unuseful for thermal history modelling and reduce experimental cost. For arbitrary combination of different low-T thermochronological data, each low-T thermochronological method is independent incompletely and the equivalent p-value is used to be the identical evaluation indicator in the inverse process.

The usefulness of the code is demonstrated by modelling thermal history of existing low-T thermochronological data in the areas of Dabie Mountain, Ahimanawa Range and Southern Alps.

**Keywords**: Low-T Thermo; Thermal history modelling; Low temperature thermochronology

# 1 Introduction

Low-T thermochronology (including apatite and zircon fission track analysis, (U–Th)/He dating and $^{40}Ar/^{39}Ar$ etc.) is a widely used tool for investigating tectonics and surface processes, providing quantitative thermal history information of a rock sample. Over the years, a number of increasingly excellent programs have been developed to extract thermal history from low-T thermochronological data (e.g., Corrigan, 1991; Gallagher, 1995; Willett, 1997; Ketcham et al., 2000; Ketcham et al., 2005; Gallagher, 2012). Except using apatite fission track (AFT) to inverse thermal history, some different low-T thermochronological combinations have been used to jointly inverse thermal history (Ketcham, 2005). For example, combining (U–Th)/He and fission track to inverse thermal history, instead of using (U–Th)/He age to be a constraint box and only inverse by fission track. However, the combinations modes are very limited and the numbers of combinations modes are very few.

Here, a new program is presented to model thermal history by low-T thermochronology. In this program, eight kinds of low-T thermochronological data can be combined arbitrarily to jointly inverse thermal history. These eight kinds of data include AFT age, AFT confined length, zircon fission-track (ZFT) age, apatite (U–Th)/He age (AHe), zircon (U–Th)/He (ZHe) age, mica Ar-Ar, optically stimulated luminescence (OSL) age of bedrock quartz and vitrinite reflectance ($R_o$). AFT age and confined length are decoupled into two completely independent data (Ding, 2017). In addition, mica Ar-Ar and bedrock quartz OSL age are added to model thermal history. Therefore, there are 247 kinds of combination modes in theory (except the eight single methods modelling) without considering the local geological situation. These arbitrary combinations are helpful to model thermal history with the "incomplete" data set regarded to be unuseful for thermal history modelling and reduce experimental cost.

The program has been designed for ease-of-use by non-specialists. The program is available by contacting the author and is free for academic, non-profit research. The operating instruction and update information is on the website: http://low-T.me.

## 2 Thermal history modelling method
### 2.1 Forward modelling

In the forward modelling procedure, the thermal history is firstly discretized into 100 evenly spaced time steps except that time step of OSL is 1 ka, then the thermal history is transferred into a modeled age or the reduced length distribution of the confined tracks.

**Apatite fission track age or confined length**

The fanning curvilinear fit annealing model of Ketcham et al. (2007) is used for C-axis projected track-lengths, assuming that $r_{mr0}$ value is 0.83. The equation for the fission-track length for C-axis projected track-lengths is from Ketcham et al. (2007). The initial C-axis projected track length is 16.62 μm (Ketcham et al., 2009). It is assumed that a minimum detectable length is 7.31μm for C-axis projected track-length (Donelick et al., 1999). The relationship of C-axis

projected mean versus standard deviation is

$\sigma_{c,mod} = 0.00615 \times l_{c,mod}^2 - 0.177194 \times l_{c,mod} + 1.829975$ which is obtained by fitting the data from Carlson et al. (1999) based on the projection model of Ketcham (2007) (Appendix: Ketcham et al., 2007). The fission-track modeled date is calculated by the cumulated track density in each time step divided by 0.893 (Ketcham et al., 2000). The conversion model from fission-track length to density presented by Donelick et al. (1999) is used here. The Kolmogorov-Smirnov (K-S) test probability, i.e. p-value (Marsaglia et al., 2003), is calculated based on the C-axis projected confined track-length distribution. The forward modelled age and AFT length distribution is statistically consistent with HeFTy (Ketcham, 2005).

**Zircon fission track age**

Annealing model formula is $r = [(6.24534 - 0.11977 \frac{\ln(t) + 314.937}{\ln(1/T) + 14.2868})^{-1/0.05721} + 1]^{-1}$ Guenthner et al. (2013). The relation $\rho = 1.25(r - 0.2)$ $(r \geq 0.36)$ (Tagami et al., 1990; Guenthner et al., 2013) is used to convert reduced length (r) to reduced density (ρ). KT06 zircon can be used as the age standard for calibration which has 10.89 μm unannealed spontaneous track length and 10.94 μm induced track length respectively (Tagami et al., 1990). Therefore, the ratio of spontaneous track length to induced track length in the standard is 0.995. Because it is closed to 1, the ratio is also can be set 1.

**Apatite and zircon (U-Th)/He**

The spherical diffusion equation (Carslaw and Jaeger 1959), Arrhenius formula, (U-Th)/He age calculation formula (Farley, 2002) and the radiation damage accumulation and annealing model for apatite (RDAAM: Flowers et al, 2009) and for zircon (ZrRDAAM: Guenthner et al., 2013) are used to represent the relationship between (U-Th)/He age and the thermal history. For (U-Th)/He, the alpha stopping distance can be defined by Farley et al. (1996) or Ketcham et al. (2011). The finite difference method (Ketcham, 2005) is used to calculate (U-Th)/He modeled age from the thermal history. The forward modeled AHe age is statistically concordant with the AHe modeled age calculated by HeFTy (Ketcham, 2005).

**Vitrinite reflectance**

There are currently three vitrinite reflectance calibrations in Low-T Thermo, the widely used EASY %Ro method of Sweeney and Burnham (1990), the "IKU" calibration described by Ritter et al. (1996), and "basin %Ro" calibration of Nielsen et al. (2016). Low-T Thermo begins the $R_o$ value calculation since the given deposition time (Stratigraphic Age) of the sample and assumes $R_o$ value formed after the deposition event. If Stratigraphic Age > modelling duration or Stratigraphic Age is not input, the modelling duration will be as Stratigraphic Age.

**Bedrock quartz OSL**

Following Herman et al. (2010), the formula describing accumulation of trapped electrons with decreasing temperature $\frac{dN}{dt} = (P - \frac{N}{A_{sat}}) - N * D_0 * e^{\frac{-E_a}{RT}}$ (Randall and Wilkins, 1945) is used. Where $N$ is the number of trapped electrons, $t$ is the time (a), $E_a$ is the activation energy (J/mol), $D_0$ is the frequency factor (m²/s) in the Arrhenian expression of diffusion coefficient, $R$ is the gas law constant (8.3145 J/(mol·K)), $T$ is the absolute temperature (K), $P$ is a filling rate (a$^{-1}$) and $A_{sat}$ is the saturation age (a). The finite difference method is used to calculate OSL modeled age from the thermal history.

**Mica Ar-Ar**

The spherical diffusion equation (Carslaw and Jaeger 1959) is used to represent the relationship between Ar-Ar age and the thermal history without considering pressure factor etc. The sphere radius is calculated using the closure temperature expression given by Dodson (1973):

$$T_c = \frac{E_a/R}{\ln(\frac{-ART_c^2 D_0/a^2}{E_a dT/dt})}$$ , where $a$ is the radius for sphere (μm), A is a geometric factor (55 for the sphere), and $T_c$ is the closure temperature (K), $dT/dt$ is the cooling rate (℃/Ma). After mica Ar-Ar closure temperature at the cooling rate is assumed, the sphere radius can be calculated. Then the finite difference method is used to solve the diffusion equation. In Low-T Thermo v1.0, after the widely acceptable $T_c$ range at 10 ℃/Ma is given, $a$ range can be calculated. Therefore, at a given thermal history, the modeled result is an age range.

**2.2 Inverse modelling**

In this inverse process, we used the Monte Carlo method to randomly search thermal histories (e.g., 10,000) where time-temperature points are not regularly distributed and can be randomly perturbed.

For comparison of different methods, p-value is used to be the identical factor to evaluate misfit. For apatite fission-track length, p-value is K-S test p-value. For ages, equivalent p-values are taken by assuming the grain ages have a normal distribution:

$$p\text{-}value = 1 - \int_{O-|O-M|}^{O+|O-M|} \frac{e^{-\frac{(x-O)^2}{2\sigma^2}}}{\sqrt{2\pi}\sigma} dx$$

. Where $O$ is the measured age, $M$ is the modeled age and $\sigma$ is the standard deviation of measured age. The 1σ age standard deviation is equivalent to K-S test p-value 0.32. When different methods are combined to model the thermal history, the minimum p-value of different methods is used to be the result.

The mean value of thermal histories is selected within a threshold equivalent p-value as the result of the thermal modelling. The threshold equivalent p-value, i.e., the acceptable goodness of fit (GOF) can be set as 0.5, 0.32, 0.05 etc. according to the data set. Because the time-step size is variable, each selected thermal history is subdivided into 100 evenly spaced time steps to calculate the mean temperature value at each time node.

For multiple grains (U–Th)/He ages, "Using multiple ages" is used to inverse each grain one by one. "Using mean age" is used to inverse using the mean age of multiple grains considering the error propagation. When "Using mean age" is used, each grain Helium content and its error are calculated from the $^{235}U$, $^{238}U$ and $^{232}Th$ content and their errors by age calculation formula (Farley, 2002). Then the mean $^{235}U$, $^{238}U$, $^{232}Th$ and He content and their errors of multiple grains are used to calculate mean (U–Th)/He age and its error by age calculation formula (Farley, 2002).

Because OSL method needs more accurate modeling than other methods between 0 and 0.5–1 million years, at least one constraint box between 0~1 million years is needed. In addition, when Bedrock Quartz OSL is selected, the order of magnitude of "Duration" should be single digit (<10 Ma) as far as possible and smaller is better.

Mica Ar-Ar method uses the modeled age range as modeled result and the measured age within the acceptable GOF is also an range. Therefore, if there is a cross between the two ranges, the assumed thermal history is thought to be acceptable.

## 3 General workflow

Low-T Thermo is written for the Microsoft Windows operating system, mainly using Mathematica. Therefore, it needs to install Mathematica 9 or 8 firstly and copy the Wolfram.NETLink.dll into the same directory with main program, Low.exe. In addition, Microsoft .NET Framework 2.0 is also needed to install.

The workflow within Low-T Thermo is as shown in Fig. 1. Different low-T thermochronological methods are firstly need to be selected, then the corresponding data (e.g., age) can be easily to copy and paste for input. Some parameters are needed to be input such as the surface temperature at sea level, the elevation, the atmospheric lapse, the maximum temperature value for the temperature axis, the maximum time value for the time axis, the number of paths tried during inversion, the acceptable GOF, and the boxes for geological constraint.

After modelling, modeled thermal history, modeled ages or values of different low-T thermochronological methods selected are displayed. "Modeled AFT length distribution" form is used to show the AFT confined track length distribution. "Modeled He age distribution" form is used to show the He age distribution with effective U concentration (eU) according the radiation damage accumulation and annealing model for apatite (RDAAM: Flowers et al, 2009) and for zircon (ZrRDAAM: Guenthner et al., 2013).

All output graphs and data can be exported to working directory. The graphs are saved as PDF-format files which can be edited in a variety of graphics packages, such as Adobe Illustrator, CorelDRAW.

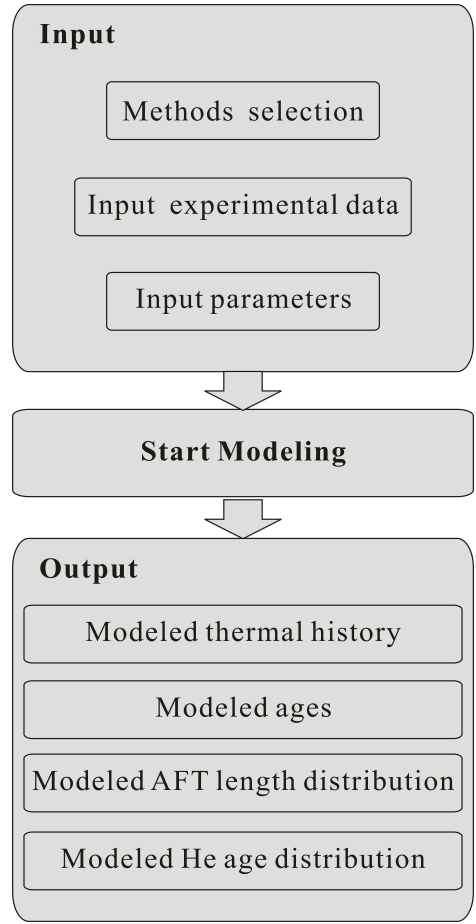

Fig 1. Flow chart describing various input options and parameters of Low-T Thermo as well as output fields.

## 3 Real examples

Examples are now presented to test the different combinations. DB40 is from Reiners et al. (2003) and Zhou et al. (2003), JR11-08 from Jiao et al. (2014), and the mean age value of samples in Southern Alps (MSA) (Herman et al., 2009; 2010) are chosen as test cases. The present day surface temperature $T_s = T_{s0} - \beta \times h$, where $T_{s0}$ is the surface temperature at sea level, $\beta$ is the atmospheric lapse rate and $h$ is the elevation.

**Example 1**

DB40 is located in Tiantangzhai peak, Dabie Mountain, China, which experienced rapid exhumation in the Late Cretaceous. For modelling test, AFT age, ZFT age and biotite $^{40}Ar/^{39}Ar$ age of DB40 are used to model thermal history. The age data is shown in Table 1. $T_{s0}$ is 15 ℃. $\beta$ is assumed as 6 ℃/km. The ratio of ZFT spontaneous track length to induced track length in the standard is 1. The closure temperature of biotite $^{40}Ar/^{39}Ar$ age is 350–400 ℃ for 10 ℃/Ma cooling rate (Grove and Harrison, 1996; Harrison et al., 1985) at . Activation energy is 210 kJ/mol, frequency factor is 0.40 cm$^2$/s (Grove and Harrison, 1996).

In this inverse process, 10,000 thermal histories were used to randomly search. The thermal history modelling result is as shown in Fig. 2. The resulting models are very similar to those determined by Zhou et al. (2003) based on both AFT age and confined length using AFTSolve software (Ketcham et al., 2000).

Table 1. The age data of DB40

| Sample Name | Elevation (m) | AFT age (Ma) | 1σ (Ma) | ZFT age (Ma) | 1σ (Ma) | $^{40}Ar/^{39}Ar$ age (Ma) | 1σ (Ma) |
|---|---|---|---|---|---|---|---|
| DB40 | 1729 | 69.5 | 3.8 | 109 | 9 | 121.5 | 2.3 |

Note: AFT age and ZFT age of DB40 are from Reiners et al. (2003) and Zhou et al. (2003), and biotite $^{40}Ar/^{39}Ar$ age is plateau age from Chen et al. (1995).

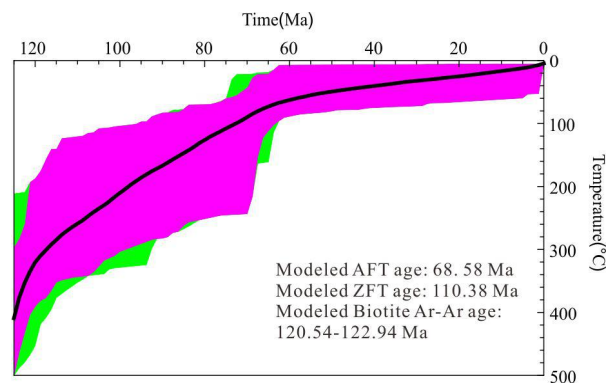

Fig 2. Model result of AFT age + ZFT age + biotite $^{40}Ar/^{39}Ar$ age of DB40 in Dabie Mountain, China. 10,000 thermal histories are used for the Monte Carlo random search. The green range is more than Acceptable Fit. The pink range is more than high GOF (≥0.5). The black line, the mean history of all paths with high GOF (≥0.5) is as modeled result. The corresponding modeled age results are also shown in textboxes.

# Example 2

JR12-14 is from Ahimanawa Range, central North Island, New Zealand, where the basement rocks were exhumed to shallow depths of the crust in the Early Cretaceous then followed by reheating before a second exhumation to shallow depths of the crust again. For modelling test, AFT confined length, AHe age of JR12-14 are used to model thermal history. The age data is shown in Table 2. $T_{s0}$ is 11 ℃. $\beta$ is assumed as 5 ℃/km.

In this inverse process, 100,000 thermal histories were used to randomly search. The thermal history modelling result is as shown in Fig. 3a. The resulting models are similar to those of Jiao et al. (2014) by QtQt software (Gallagher, 2012) based on AFT (both age and confined length) and AHe age from a vertical profile. However, it must be mentioned that QtQt doesn't need the constraint box, the constraint box is needed by Low-T Thermo for reheating modelling. Modeled He age distribution (Fig. 3b) shows the He age distributions of two grains with eU.

Table 2. AHe data and AFT length data of JR12-14

| Sample name | Elevation (m) | Grain No. | U (ppm) | Th (ppm) | $F_T$ | AHe age (Ma) | 1σ (Ma) | AFT length |
|---|---|---|---|---|---|---|---|---|
| JR12-14 | 916 | 1 | 6.8 | 29.5 | 0.67 | 36.5 | 2.7 | - |
|  |  | 2 | 7.0 | 9.9 | 0.71 | 30.6 | 2.4 |  |

Note: JR12-14 is from Jiao et al. (2014).

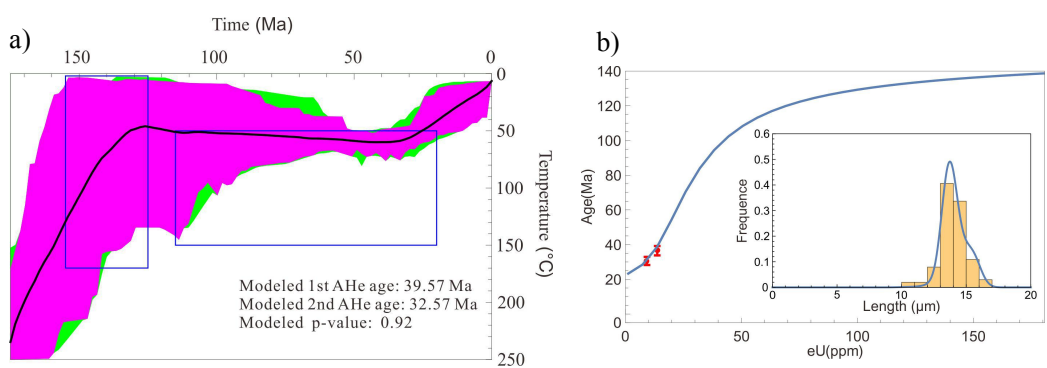

Fig. 3 The thermal history modelling using different AHe grain-age data and confined track lengths (AHe age + AFT length) of JR12-14 in Ahimanawa Range, New Zealand. a) JR12-14 thermal history modelling based on combining three AHe grain ages and the AFT confined track lengths. All the thermal histories have ≥0.05 p-values. 100,000 thermal histories are used for the Monte Carlo random search. The minimum equivalent p-values is taken as the evaluating parameter. The green range is more than Acceptable Fit. The pink range is more than high GOF (≥0.5). The black line, the mean history of all paths with high GOF (≥0.5) is as modeled result. The corresponding modeled age results are also shown in textboxes. b) The corresponding modeled He age distribution with eU according RDAAM (Flowers et al, 2009) and AFT length distribution.

## Example 3

The samples in the central Southern Alps of New Zealand experienced cyclic glaciations that were initiated in the Pliocene around 2.5±0.1 Ma (Suggate, 1990) and an extreme exhumation happened in 0.1 Ma (Herman et al., 2010). For modelling test, AHe age, ZHe age and Bedrock Quartz OSL of MSA are used to model thermal history. The age data is shown in Table 3. Ts0 and β is same as Example 2.

In this inverse process, 100,000 thermal histories were used to randomly search. A box from 0.1 to 0.5 Ma is used. The thermal history modelling result is as shown in Fig. 4. The resulting models are very concordant with that estimated by Herman et al. (2010).

Table 3. AHe, ZHe and Bedrock Quartz OSL age of MSA

| Sample name | Elevation (m) | U (ppm) | Th (ppm) | $F_T$ | AHe age (Ma) | 1σ (Ma) | ZHe age (Ma) | 1σ (Ma) | total dose rate (mGy/a) | Equivalent dose (Gy) | OSL age (ka) | 1σ (ka) |
|---|---|---|---|---|---|---|---|---|---|---|---|---|
| MSA | 1454 | 5.08 | 3.14 | 0.75 | 0.47 | 0.11 | | | 3.27 | 251.23 | 81.62 | 13.46 |
| | | 315.98 | 74.13 | 0.79 | | | 1.58 | 0.06 | | | | |

Note: All of the data are the mean value of samples in Southern Alps (Herman et al., 2009; 2010)

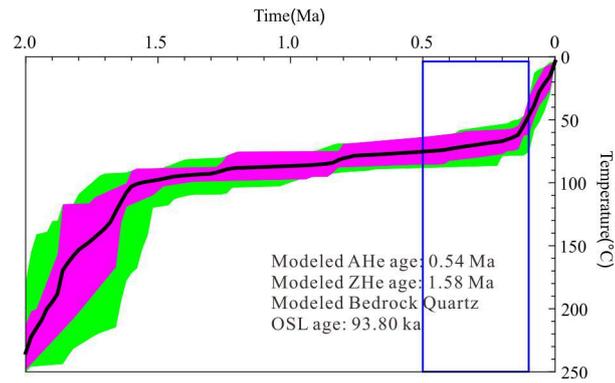

Fig 4. Model result of mean AHe age + ZHe age + Bedrock Quartz OSL age of the samples in Southern Alps, New Zealand. 100,000 thermal histories are used for the Monte Carlo random search. The green range is more than Acceptable Fit. The pink range is more than high GOF (≥0.5). The black line, the mean history of all paths with high GOF (≥0.5) is as modeled result. The corresponding modeled age results are also shown in textboxes.

## 4. Conclusions

A new, user-friendly, code (called Low-T Thermo) has been developed to combine low-T thermochronological data arbitrarily to model thermal history. These low-T thermochronological data include AFT age, AFT confined length, ZFT age, AHe age, ZHe age, $R_o$, mica Ar-Ar age and bedrock quartz OSL age. They are independent incompletely. Although different combination has different advantage in thermal history modelling because every kind of low-T thermochronological data has its own advantage in constraining thermal evolution, these arbitrary combinations are useful to model thermal history with the "incomplete" low-T thermochronological data set regarded to be unuseful for thermal history modelling in the past and reduce experimental cost.